\documentstyle[12pt]{article}
\begin{document}
\pagestyle{empty}
\baselineskip=17pt
\noindent November 1994 \\
Submitted to Phys.\ Rev.\ E
\bigskip
\bigskip
\begin{center}
\begin{large}
{\bf Exact Exponent $\lambda$ of the Autocorrelation Function for a
Soluble Model of Coarsening}\\
\end{large}
\bigskip
A. J. Bray$^1$ and B. Derrida$^{2,3}$ \\
\bigskip
$^1$Department of Physics and Astronomy, The University,
Manchester M13 9PL, UK. \\
$^2$Laboratoire de Physique Statistique, Ecole Normale Sup\'{e}rieure,
24 Rue Lhomond, 75231 Paris cedex 05, France. \\
$^3$Service de   Physique Th\'{e}orique, CE Saclay, 91191 Gif sur Yvette,
France. \\
\bigskip
\bigskip
{\bf Abstract} \\
\end{center}
\bigskip
The exponent $\lambda$ that describes the decay of the autocorrelation
function $A(t)$ in a phase ordering system, $A(t) \sim L^{-(d-\lambda)}$,
where $d$ is the dimension and $L$ the characteristic length scale at time
$t$, is calculated exactly for the time-dependent Ginzburg-Landau equation
in $d=1$. We find $\lambda = 0.399\,383\,5\ldots$. We also show explicitly
that a small bias of positive domains over negative gives a magnetization
which grows in time as $M(t) \sim L^\mu$ and prove that for the $1d$
Ginzburg-Landau equation, $\mu=\lambda$, exemplifying a general result.

\newpage
\pagestyle{plain}
\pagenumbering{arabic}

The field of phase ordering kinetics has seen a number of new developments
in recent years \cite{BrayRev}. In particular the values of the growth
exponents $z$, which describe the time-dependence of the characteristic
scale $L(t)$ via $L \sim t^{1/z}$, are known exactly for most models with
purely dissipative dynamics \cite{BrayRev,Exponents}. For systems with
short-range interactions and dynamics which are either nonconserved or
obey a local conservation law, the exponent $z$ is usually a
dimension-independent integer \cite{Exponents}. Recently, however, it has
been realized that for nonconserved dynamics the description of two-time
correlations requires a new exponent, whose dependence on the spatial
dimension $d$ and on the symmetry of the order parameter is nontrivial
\cite{FH88,Newman}.

The exponent $\lambda$ can be defined in terms of the general two-point
correlation function
$C(r;t_1,t_2) = \langle \phi({\bf x},t_1)\,\phi({\bf x}+{\bf r},t_2)\rangle$,
where $\phi$ is the order parameter field. In the scaling regime, this is
expected to have the scaling form $C(r;t_1,t_2) = f(r/L_1,r/L_2)$, where
$L_1$, $L_2$ are the characteristic length scales at times $t_1$ and $t_2$
\cite{Furukawa,Rutenberg}. In the limit of well-separated times,
$L_2 \gg L_1$, one anticipates \cite{Furukawa} the power-law form
$C(r;t_1,t_2) \sim (L_1/L_2)^{d-\lambda} f(r/L_2)$, defining the exponent
$\lambda$. An especially simple case is where we take $r=0$, and the initial
time $t_1=0$. Then the general form reduces to
\begin{equation}
A(t) \equiv C(0;0,t) \sim [\xi_0/L(t)]^{d-\lambda}\ ,
\label{AUTO}
\end{equation}
where $\xi_0$ is some fixed length related
to the initial conditions. The `autocorrelation function' $A(t)$ has been
measured in simulations of $O(n)$ models for various spatial dimensions
$d$ \cite{FH88,Sims}, and in experiments on twisted nematic liquid crystals
films \cite{Yurke}, and the exponent $\lambda$ deduced. It generally has a
nontrivial value.

There are a few analytical results for $\lambda$ -- the
nonconserved $O(n)$ model for $n=\infty$ ($\lambda=d/2$ \cite{Newman}),
and the $d=1$ Glauber model ($\lambda=0$ \cite{Bray}), while for
nonconserved scalar fields in $d=2$ Fisher and Huse \cite{FH88} have
conjectured that $\lambda=3/4$ exactly. In general, however, $\lambda$
appears to be a nontrivial exponent associated with ordering dynamics,
although it is known to satisfy the bound (in our notation)
$\lambda \le d/2$ for nonconserved dynamics \cite{FH88,Yeung}.

In this paper we calculate $\lambda$ exactly for a soluble model
corresponding to the late-time, zero-temperature coarsening dynamics of
the time-dependent Ginzburg-Landau (TDGL) equation for a scalar field in
$d=1$. The equation of motion is $\partial_t \phi = \partial_x^2 \phi
- dV/d\phi$, where $V(\phi)$ is a symmetric, double-well potential with
minima at $\phi=\pm 1$ (e.g.\ $V(\phi) = (1-\phi^2)^2$).
At late times, when the mean separation $L$ of domain walls is large
compared to their intrinsic width $\xi$ ($=[V''(1)]^{-1/2}$), the walls
only interact weakly, through the exponential tails of the wall profile
function. Then the dynamics is very simple \cite{Kawasaki,RB,BDG}. The
closest pair of walls move together and annihilate, while the other walls
hardly move at all, and the system coarsens by successively eliminating
the smallest domains. It is found that the distribution of domain sizes
$l$ approaches a scaling form, $P(l) = L^{-1}f(l/L)$. The scaling function
$f(x)$ can be exactly calculated \cite{Kawasaki,RB,BDG}.

In an earlier work \cite{BDG}, we have shown that there is a nontrivial
exponent associated with the fraction of the line that has never been
traversed by a domain wall (i.e.\ the fraction of the line where the order
parameter $\phi$ has never changed its sign \cite{DBG}).
 This fraction decays as $L^{-(1-\beta)}$,
with $\beta = 0.824\,924\,12\dots$. Here we show that the  approach
developed in \cite{BDG}
can be generalised to calculate $\lambda$ for this model. The result is
$\lambda = 0.399\,383\,5\ldots$. A recent simulation of the same
model \cite{MH} gave the estimate $\lambda = 0.43 \pm 0.01$,
which, we think, is in reasonable agreement with our exact result,
given that the extrapolation to large $L$ was not straightforward.

The exponent $\lambda$ can also be obtained from the rate at which a small
initial bias in the order parameter grows with time \cite{Kiss},
$\langle \phi \rangle \sim L^\lambda$. We demonstrate this explicitly
within the present model  in the second part of this work.

The calculation of the autocorrelation exponent $\lambda$ follows closely
 that presented in reference \cite{BDG}.
One starts with random intervals on the line. Each interval $I$ is
characterised by its length $l(I)$ and by  its  overlap $q(I)$ with its
initial condition (initially $q(I)=l(I)$ for all $I$). At each iteration
step, the smallest interval $I_{\rm min}$
is removed (i.e.\ the field $\phi$ is replaced by $- \phi$ in this interval).
So three intervals (the smallest interval $I_{\rm min}$ and
its two neighbors $I_1$ and $I_2$) are  replaced  by a single interval $I$.
The length and the overlap of the new interval $I$ are given by
\begin{eqnarray}
l(I) & = & l(I_1) +l(I_{\rm min}) + l(I_2)\ , \\
q(I) & = & q(I_1) +  q(I_2) -q(I_{\rm min})\ .
\end{eqnarray}
 Then the average length $L$ of domains and   the autocorrelation function
$A$ are given by
\begin{eqnarray}
L & = & \sum_I l(I) / \sum_I 1  \ ,  \nonumber \\
 A & = & \sum_I q(I) / \sum_I l(I) \ .
\label{sum}
\end{eqnarray}
where the sums are over all the intervals  $I$ present in the system.

The argument showing that no correlations develop if none are present
initially was given earlier \cite{BDG} and  the calculation is then very
similar to that for the evaluation of the exponent $\beta$.
We take, for simplicity, the  lengths of the intervals  to be integers
and  $i_0$ to be the minimal length in the system. We  also assume that
the total  number  $N$ of intervals is very large. We call $n_i$ the
number of intervals of length $i$  and $q_i$  the average overlap of the
intervals of length $i$. At the beginning, $q_i= i$.

We denote with a prime the values of these quantities after all
the $n_{i_0}$ intervals of length $i_0$ have been eliminated, so that
the minimal length has become $i_0 + 1$.
Then the time evolution is given by (compare equation (2) of \cite{BDG})
\begin{eqnarray}
N' & = & N- 2 n_{i_0} \nonumber \\
n_i' & = & n_i(1 - {2 n_{i_0} \over N}) + n_{i_0} \sum_{j=i_0}^{i-2i_0}
{ n_{j} \over N}  \ { n_{i-j-i_0} \over N} \nonumber \\
n_i' q_i' & = & n_i q_i(1 - {2 n_{i_0} \over N}) +
n_{i_0} \sum_{j=i_0}^{i-2i_0}{ n_{j}  \over N} \
{ n_{i-j-i_0} \over N}  \ ( q_j + q_{i-j-i_0} -q_{i_0})\ .
\label{EQ:ITERATE}
\end{eqnarray}
This is only valid under the condition that $n_{i_0} \ll N$ which is indeed
valid when $i_0$ becomes large and as long as the system consists of a large
number of intervals.

We assume that after many iterations, i.e. when $i_0$ becomes large,
a scaling limit is reached where
\begin{eqnarray}
n_i & = & \frac{N}{i_0} f \left( {i \over i_0} \right) \nonumber \\
n_i q_i & = & N (i_0)^{\lambda-1} g \left( {i \over i_0} \right)\ ,
\label{SCALING}
\end{eqnarray}
where $\lambda$ is the exponent  we want to calculate (\ref{sum}).
Because $i_0$ is so large,  we can consider $x=i/i_0$ as a continuous
variable. This gives
\begin{eqnarray}
n_i' & = & \frac{N'}{i_0+1} f \left( {i \over i_0+1} \right) =
{N \over i_0} \left[f(x) - {2 \over i_0} f(1)f(x) - {1 \over i_0} f(x)
-{1 \over i_0}  x f'(x) \right] \nonumber \\
n_i'q_i' & = & N' (i_0+1)^{\lambda-1} g \left( {i \over i_0+1} \right)  \\
 & = & {N i_0^{\lambda-1}} \left[g(x)  - {2 \over i_0} f(1)g(x)
+ {\lambda -1 \over i_0} g(x) -{1 \over i_0}  x g'(x)  \right]\ .
\end{eqnarray}
Inserting these expressions in the time evolution equations
(\ref{EQ:ITERATE})
 gives
\begin{eqnarray}
i_0 {\partial f \over \partial i_0} & = &
f(x) + x f'(x) + \theta(x-3) f(1) \int_1^{x-2} dy \  f(y) f(x-y-1)
 \nonumber \\
i_0 {\partial g \over \partial i_0} & = &
(1 -\lambda)g(x) + x g'(x) + 2  \theta(x-3) f(1) \int_1^{x-2} dy
\ g(y) f(x-y-1)   \nonumber \\
 &&  -g(1) \theta(x-3) \int_1^{x-2} dy \  f(y) f(x-y-1) \ .
\label{9}
\end{eqnarray}
In (\ref{SCALING}), both $n_i$ and $n_i q_i$ are functions of $x=i/i_0$
and of $i_0$, and the partial derivatives in (\ref{9}) mean the derivative
 with respect to $i_0$, keeping $x$ fixed.
Demanding that the system is self-similar, i.e.\ that the functions
 $f(x)$ and $g(x)$ do not change with time (i.e.\ replacing the left-hand
 sides of (\ref{9}) by zero), one finds that
 the Laplace transforms
\begin{eqnarray}
\phi(p) & = & \int_1^\infty e^{-px}  \ f(x) \ dx\ , \nonumber \\
\psi(p) & = & \int_1^\infty e^{-px}  \ g(x) \ dx\ ,
\label{EQ:LAPLACE}
\end{eqnarray}
 satisfy the  following equations (where primes now indicate derivatives)
\begin{eqnarray}
\label{PHI}
-f(1) e^{-p} - p \phi'(p) + f(1) e^{-p} \phi^2(p) & = & 0 \\
 -\lambda \psi(p) - g(1) e^{-p} - p \psi'(p)
 + 2 f(1) e^{-p} \phi(p) \psi(p) -g(1) e^{-p} \phi^2(p) & = & 0\ .
\label{PSI}
\end{eqnarray}
Defining the function $h(p)$ by
\begin{equation}
h(p) =  2 f(1) \int_p^\infty {e^{-t} \over t} dt\ ,
\end{equation}
the solutions of the above equations are
\begin{eqnarray}
\label{EQ:PHI}
\phi(p) & = & \tanh [h(p)/2] \\
\psi(p) & = & g(1) \int_p^\infty (1 + \phi^2(q))  { 1-\phi^2(p) \over
 1-\phi^2(q) } \ {q^{\lambda -1} \over p^\lambda} \  e^{-q} dq\ .
\label{EQ:PSI}
\end{eqnarray}
The constants of integration implied by these forms were fixed by the
requirement that both $\phi$ and $\psi$ decay fast enough for large  $p$,
as is clear from the definitions (\ref{EQ:LAPLACE}).
So far the parameters $f(1)$, $g(1)$ and $\lambda$ are arbitrary.
We shall see that they are fixed by physical considerations.

Eq.\ (\ref{EQ:PHI}) for $\phi$, which determines the domain size
distribution, is of course identical to that obtained in previous work
\cite{Kawasaki,RB,BDG}. Eq.\ (\ref{EQ:PSI}) for  $\psi$  can be
rewritten in the more convenient form
\begin{equation}
\psi(p) =  2 g(1)  \int_p^\infty  { e^{h(q)} + e^{-h(q)} \over
 e^{h(p)} +2 + e^{-h(p)} } \
 {q^{\lambda -1} \over p^\lambda}  \ e^{-q} dq \ .
\label{CONV}
\end{equation}
It is helpful to introduce the expansion
\begin{equation}
\int_p^\infty {e^{-q} \over q} dq = -\log p
-\gamma -\sum_{n=1}^{\infty} {(-p)^n \over n \ n!}\ ,
\label{EXPANSION}
\end{equation}
where  $\gamma = -\int_0^\infty dt \  e^{-t} \  \log t
= .577\,215\,6... $ is Euler's constant.

{}From the small-$p$ expansion of (\ref{EQ:PHI}), it is easy to show that,
provided the first moment of the domain size distribution exists, one must
have $f(1)=1/2$ \cite{Kawasaki,RB,BDG}. From now on, we will consider only
this case(see \cite{DGY} for the discussion of cases where the stationary
distribution has long tails). Defining the function $r(p)$ by
\begin{equation}
r(p) = h(p)+\log p  = \int_p^\infty {e^{-q} \over q} dq + \log p\ ,
\label{r}
\end{equation}
one  obtains, using (\ref{CONV}),
\begin{equation}
\psi(p) = 2  g(1)  \int_p^\infty  { e^{r(q)}  +q^2 e^{-r(q)} \over
e^{r(p)} +2 p + p^2 e^{-r(p)} } \
{q^{\lambda -2} \over p^{\lambda-1}}  \ e^{-q} dq\ .
\end{equation}
Now $r(p)$ can be expanded in powers of $p$, using (\ref{EXPANSION}), and
so this last form makes it easier to analyse the singular behavior of
$\psi(p)$ at $p=0$. One finds that, for small $p$,
\begin{equation}
\psi(p)  = A + B p^{1-\lambda}[1+O(p)]\ ,
\label{EQ:SING}
\end{equation}
where $A=2g(1)/(1 - \lambda)$ and
\begin{eqnarray}
B & = & 2 g(1) e^{-r(0)} \left[  \int_0^\infty {q^{\lambda-1} e^{-q}
\over 1 - \lambda} (r'(q) -1) e^{r(q)} dq
+ \int_0^\infty q^\lambda e^{-q}  e^{-r(q)} dq  \right]
\label{Cond1}
 \\
& = & 2 g(1) e^{\gamma}(1-\lambda)^{-1} \int_0^\infty  q^{\lambda-2} e^{-q}
 \left[ (1-q-e^{-q}) e^{r(q)} +q^2 (1 - \lambda) e^{-r(q)} \right] dq \ .
\nonumber
\end{eqnarray}

Now compare (\ref{EQ:SING}) with a direct expansion of (\ref{EQ:LAPLACE}),
namely $\psi(p) = \int_1^\infty dx g(x)(1-px +O(p^2))$. If the function
$g(x)$ is to have a finite first moment then we must have $B=0$ in
(\ref{EQ:SING}). This condition determines $\lambda$ as
\begin{equation}
\lambda = .399\,383\,5\ldots\ .
\label{lambda}
\end{equation}
{}From numerical simulations of the same model, Majumdar and Huse \cite{MH}
found the power-law decay $A(t) \sim L^{-\bar{\lambda}}$, with
$\bar{\lambda} = 0.57 \pm 0.01$, corresponding to
$\lambda \equiv d-\bar{\lambda} = 0.43 \pm .01$. There were, however,
large corrections to scaling in their numerical data, which we think are
the origin of the disagreement between their numerical estimate and our
exact result.

As in \cite{BDG}, one can show that $B \neq 0$ would correspond to a power
law decay in $g(x)$ and that such a power law  cannot be produced if it is
not present in the initial condition. Note that $g(1)$ cannot be determined
as one can always multiply all the $q_i$ by a constant without changing our
results.

For the remainder of this paper we will look at a related quantity,
the growth of an initially small bias in the order parameter, and show
that the bias grows as $L^\mu$ as the system coarsens (while the bias
remains small). Furthermore, we will show explicitly that $\mu=\lambda$
for this model, exemplifying a general result \cite{Kiss}.

Consider a sequence of positive and negative domains on a line.
We call $n_i$ ($m_i$) the number of positive (negative) domains of
length $i$. The total number $N$  of positive domains
is of course equal to the total number of negative domains,
$ N = \sum_i n_i = \sum_i m_i $. When the domains of size $i_0$ are
removed, the new values of $n_i$, $m_i$ and $N$ are given by
\begin{eqnarray}
n_i' & = & \left( 1 - {2 m_{i_0} \over N} \right)
+ m_{i_0} \sum_{j=i_0}^{i-2i_0} {n_j  \ n_{i-j-i_0} \over N^2} \nonumber \\
m_i' & = & \left( 1 - {2 n_{i_0} \over N} \right)
+ n_{i_0} \sum_{j=i_0}^{i-2i_0} {m_j \  m_{i-j-i_0} \over N^2} \nonumber \\
N' & = & N- n_{i_0} - m_{i_0}\ .
\end{eqnarray}
Let us write forms for $n_i$ and $m_i$ analogous to the first of equations
(\ref{SCALING}):
\begin{equation}
n_i = {N \over i_0} f_1\left( {i \over i_0} \right)\ ,\ \ \ \ \
m_i = {N \over i_0} f_2\left( {i \over i_0} \right)\ .
\label{PSEUDOSCALING}
\end{equation}
Then one has
\begin{equation}
n_i' = {N- n_{i_0} - m_{i_0} \over i_0 +1} f_1 \left( {i \over i_0 +1}
\right)\ ,
\end{equation}
which gives, for $i_0$ large (when $x=i/i_0$ can be treated as a continuous
variable),
\begin{equation}
n_i' = {N \over i_0} \left[ f_1(x) +{1 \over i_0}
\left\{ - f_1(1) f_1(x) - f_2(1)f_1(x) - f_1(x) - x f_1'(x) \right\}
\right]\ ,
\label{ASYMMETRIC}
\end{equation}
and a similar expression for $m_i'$.

 Inserting the
forms (\ref{PSEUDOSCALING}) into (\ref{ASYMMETRIC}) gives coupled evolution
equations for $f_1$ and $f_2$:
\begin{eqnarray}
 i_0 { \partial f_1(x) \over \partial i_0} & = &
 [f_1(1)-f_2(1)]f_1(x) +f_1(x) + x f_1'(x) \nonumber \\
 & & + \theta(x-3) f_2(1) \int_1^{x-2} dy \  f_1(y) f_1(x-y-1)\ ,
\end{eqnarray}
and a second equation obtained by interchanging the subscripts `1' and `2'.
Note that the derivatives on the left-hand sides are with respect to the
(implicit) second argument $i_0$. Introducing the Laplace transforms with
respect to the first argument,
\begin{equation}
\psi_n(p)  =  \int_1^\infty f_n(x) e^{-px} dx\ ,\ \ \ \ \ (n=1,2)
\label{PSILAPLACE}
\end{equation}
one finds that their evolution is given by
\begin{equation}
i_0 { \partial \psi_1(p) \over \partial i_0}
= [f_1(1)-f_2(1)]\psi_1(p)  -p \psi_1'(p)  -
f_1(1) e^{-p} + f_2(1) e^{-p} \psi_1^2(p)\ ,
\label{BIAS}
\end{equation}
and a second equation with `1' and `2' interchanged.

So far this is completely general. The basic idea is to perform a
linear stability analysis around the `symmetric' solution
$\psi_1(p)=\psi_2(p)=\phi(p)$, where $\phi(p)$ satisfies (\ref{PHI})
with $f(1)=1/2$, in order to determine the rate at which a small
perturbation will grow. We therefore take $\psi_1(p)$  and  $\psi_2(p)$
to have the forms
\begin{equation}
\psi_n(p) = \phi(p) \pm \epsilon \ \sigma(p)\ ,
\label{SIGMA}
\end{equation}
with
\begin{equation}
f_n(1) = {1 \over 2} \pm \epsilon  \ a\ ,
\label{a}
\end{equation}
with $\epsilon$ small and the $+$ ($-$) sign corresponding to $n=1$
($n=2$). If the bias represented by the terms in $\epsilon$ is a relevant
perturbation, $\sigma(p)$ will grow under iteration:
$\sigma \sim (i_0)^\mu$ with $\mu>0$ (and similarly, $a \sim (i_0)^\mu$
in (\ref{a})). Subtracting from (\ref{BIAS}) its counterpart with `1' and
`2' interchanged, and putting $i_0\,\partial\sigma(p)/\partial i_0 =
\mu\sigma(p)$, yields the eigenvalue equation
\begin{equation}
\mu\sigma = 2 a \phi - p \sigma' -a e^{-p} - a e^{-p} \phi^2
+ e^{-p} \phi \sigma\ ,
\end{equation}
with solution
\begin{equation}
\sigma(p) = a \int_p^\infty  {\phi^2(q) e^{-q} + e^{-q}
- 2 \phi(q) \over q} \left( q \over p \right)^\mu {1 - \phi^2(p)
\over 1 - \phi^2(q)} dq\ .
\end{equation}
The integration constant was fixed as before by the requirement that
$\sigma(p)$ decrease as $\exp(-p)/p$ for large $p$, which follows from
(\ref{PSILAPLACE}), (\ref{SIGMA}) and (\ref{a}). Demanding once more that
$\sigma(p)$ be regular at $p=0$ (so that the first moments of $f_1(x)$
and $f_2(x)$ exist) yields the following equation for $\mu$:
\begin{equation}
\int_0^\infty dq  \ [ e^{-q} \phi^2(q) + e^{-q} - 2 \phi(q) ]
\  { q^{\mu-1} \over 1 - \phi^2(q) } =0\ .
\end{equation}
Using $\phi(q) = (e^{r(q)} - q)/(e^{r(q)} +q)$, which follows
from (\ref{EQ:PHI}) and (\ref{r}), gives the condition
\begin{equation}
 \int_0^\infty dq  \ [\  (e^{-q} -1 )\  e^{r(q)}
 + q^2 (e^{-q} +1)\  e^{-r(q)}\  ] q^{\mu-2} =0
\label{Cond2}
\end{equation}
for $\mu$, with solution $\mu \simeq .399\,38...$.
Comparison with (\ref{lambda}) suggests that $\mu=\lambda$. In fact,
using integration by parts one can show that condition (\ref{Cond2})
for $\mu$ is identical to (\ref{Cond1}) (with $B=0$) for $\lambda$,
and so $\mu=\lambda$ exactly.

The result $\mu=\lambda$ is, in fact, quite general. For TDGL dynamics,
it has been discussed elsewhere \cite{Kiss}.  Let us derive it for any
kind of dynamics of an Ising model. Consider a system of $N$ Ising spins
in dimension $d$. We call $P(\,\{S_i(t)\}\,|\,\{S_i(0)\}\,)$ the
probability of findingthe system in the spin configuration $\{S_i(t)\}$
at time $t$ given that it was in configuration $\{S_i(0)\}$ at time $0$.
We assume that the system evolves in a zero magnetic field and that the
dynamics preserves the $\pm$ symmetry, namely
$P(\,\{S_i(t)\,\}|\,\{S_i(0)\}\,) = P(\,\{-S_i(t)\}\,|\,\{- S_i(0)\}\,)$.

Suppose that one starts with an initial condition
$\{S_i(0)\}$ chosen  completely  at random, then the
correlation $\langle S_i(t) S_j(0) \rangle $ is given by
\begin{equation}
\langle S_i(t) S_j(0) \rangle  = {1 \over 2^N} \sum_{\{S(t)\}}
\sum_{\{S(0)\}}S_i(t)S_j(0)\,P(\,\{S_i(t)\}\,|\,\{S_i(0)\}\,)\ .
\end{equation}
where $\sum_{\{S(t)\}}$ indicates a sum over the $2^N$ configurations at
time $t$.

Suppose on the other hand that one starts with a weakly magnetized
initial condition, i.e.\ the initial configuration $\{S_i(0)\}$ is chosen
with probability
$$
Q(\{S_i(0)\}) = \prod_{i=1}^N {1 + m(0) S_i(0) \over 2}
  \simeq {1 + m(0) \sum_j S_j(0) \over 2^N } $$  when $m(0)$
is infinitesimal. Then the magnetization $m(t)$ per spin at time $t$ is a
function of $m(0)$, and to first order in powers of $m(0)$ one has
\begin{eqnarray}
m(t) & = & \sum_{\{S(t)\}} \sum_{\{S(0)\}}
P(\,\{S_i(t)\}\,|\,\{S_i(0)\}\,)\,Q(\{S_i(0)\})\,
{\sum_j S_j(t) \over N} \nonumber \\
 &\simeq &
 m(0)\,{\sum_i \sum_j \langle S_i(t)  S_j(0) \rangle \over N}\ .
\end{eqnarray}
Therefore if one assumes that due to some coarsening
phenomenon the two-point function scales as
$$
\langle S_i(0) S_j(t) \rangle \simeq L^{-(d-\lambda)} f({R_{ij} \over L})
$$
where $R_{ij}$ is the distance between sites $i$ and $j$,
one finds that
$$m(t) \simeq L^\lambda m(0) \int d^d R f(R)$$ which means that the
magnetization and the autocorrelation  exponents are the same.

To summarise, we have derived a non-trivial value for the exponent
$\lambda$ within an exactly soluble model, and shown explicitly that the
growth of an initial bias in the order parameter is controlled by the
same exponent.

We thank the Isaac Newton Institute, Cambridge, where this work was
carried out, for its hospitality.


\newpage

\begin{thebibliography}{99}
\bibitem{BrayRev} For a recent review see A. J. Bray, Advances in Physics,
to appear.
\bibitem{Exponents} I. M. Lifshitz, Zh.\ Theor.\ Fiz. {\bf 42}, 1354
(1962) [Sov.\ Fiz.\ JETP {\bf 15}, 939 (1962)]; S. M. Allen and J. W. Cahn,
Acta.\ Metall.\ {\bf 27}, 1085 (1979); I. M. Lifshitz and V. V. Slyozov,
J. Chem.\ Phys.\ Solids {\bf 19}, 35 (1961); A. J. Bray, Phys.\ Rev.\ Lett.
{\bf 62}, 2841 (1989); A. J. Bray and A. D. Rutenberg, Phys.\ Rev.\ E
{\bf 49}, R27 (1994).
\bibitem{FH88} D. S. Fisher and D. A. Huse, Phys.\ Rev.\ B {\bf 38}, 373
(1988). Note that our exponent $\bar{\lambda} \equiv d-\lambda$ is called
$\lambda$ in this paper.
\bibitem{Newman} T. J. Newman and A. J. Bray, J. Phys.\ A {\bf 23}, 4491
(1990); J. G. Kissner and A. J. Bray, J. Phys.\ A {\bf 26}, 1571 (1993).
\bibitem{Furukawa} H. Furukawa, J. Phys.\ Soc.\ Jpn.\ {\bf 58}, 216 (1989);
Phys.\ Rev.\ B {\bf 40}, 2341 (1989).
\bibitem{Rutenberg} This generalized scaling assumption does not always
hold however. An explicit counterexample is the $O(2)$ model for $d=1$:
see A. J. Bray, reference \cite{BrayRev}; T. J. Newman et al.,
reference \cite{Sims}; A. D. Rutenberg and A. J. Bray, preprint.
\bibitem{Sims} T. J. Newman, A. J. Bray and M. A. Moore, Phys.\ Rev.\ B
{\bf 42}, 4514 (1990); A. J. Bray and K. Humayun, J. Phys.\ A {\bf 23},
5897 (1990); K. Humayun and A. J. Bray, J. Phys.\ A {\bf 24}, 1915 (1991);
F. Liu and G. F. Mazenko, Phys.\ Rev.\ B {\bf 44}, 9185 (1991).
\bibitem{Yurke} N. Mason, A. N. Pargellis and B. Yurke, Phys.\ Rev.\
Lett.\ {\bf 70}, 190 (1993); A. N. Pargellis, S. Green and B. Yurke,
Phys.\ Rev.\ E {\bf 49}, 4250 (1994).
\bibitem{Bray} A. J. Bray, J. Phys.\ A {\bf 22}, L67 (1990).
\bibitem{Yeung} C. Yeung, M. Rao and R. C. Desai, preprint.
\bibitem{Kawasaki} T. Nagai and K. Kawasaki, Physica A {\bf 134}, 483
(1986).
\bibitem{RB} A. D. Rutenberg and A. J. Bray, Phys.\ Rev.\ E
{\bf 50}, 1900 (1994).
\bibitem{BDG} A. J. Bray, B. Derrida and C. Godr\`{e}che, Europhys.\
Lett.\ {\bf 27}, 175 (1994).
\bibitem{DBG}  B. Derrida, A.J. Bray and C. Godr\`{e}che,
J. Phys.\ A {\bf 27}, L357 (1994).
\bibitem{MH} S. N. Majumdar and D. A. Huse, preprint. Note that our
exponent $\bar{\lambda} \equiv d-\lambda$ is called $\lambda$ in this
paper.
\bibitem{Kiss} A. J. Bray and J. G. Kissner, J. Phys.\ A {\bf 25},
31 (1992).
\bibitem{DGY} B. Derrida, C. Godr\`{e}che and I. Yekutieli, Phys.\
Rev.\ A {\bf 44}, 6241 (1991).
\end{thebibliography}
\end{document}